# Characterizing the Human Mobility Pattern in a Large Street Network

Bin Jiang [(1)], Junjun Yin [(1)] and Sijian Zhao [(2)]

[(1)] Division of Geomatics, Department of Technology and Built Environment
University of Gävle, Sweden
Email: bin.jiang@hig.se, yinjunjun@gmail.com

[(2)] College of Resources Science and Technology, Beijing Normal University, China
Email: scanzhao@hotmail.com

**Abstract**
Previous studies demonstrated empirically that human mobility exhibits Lévy flight behaviour. However, our knowledge of the mechanisms governing this Lévy flight behaviour remains limited. Here we analyze over 72 000 people's moving trajectories, obtained from 50 taxicabs during a six-month period in a large street network, and illustrate that the human mobility pattern, or the Lévy flight behaviour, is mainly attributed to the underlying street network. In other words, the goal-directed nature of human movement has little effect on the overall traffic distribution. We further simulate the mobility of a large number of random walkers, and find that (1) the simulated random walkers can reproduce the same human mobility pattern, and (2) the simulated mobility rate of the random walkers correlates pretty well (an R square up to 0.87) with the observed human mobility rate.

## 1. Introduction

Human mobility in a street network consists of numerous flights (or jumps) from one street to another. To be specific, a person at the origin (O) with street *a* will randomly (sometimes with a priority to the most connected ones) decide to choose a connected street *b*; she walks along street *a*, before she reaches the junction of *a* and *b*. Then she will choose another connected street *c*, and so on and so forth until she has reached her destination (D) (Figure 1). Thus human mobility in a street network, a sort of network-constrained movement, is like a frog jumping from one street to another at the topological level, and like a turtle walking persistently along individual streets at the geometric level. This movement behaviour might be over simplified, because human mobility in reality tends to be purposive with destinations such as schools, offices, and homes. Despite the purposive nature of human mobility at the individual level, human movement tend to be predictable at the level of crowds in terms of how many people come to individual streets. This is the basic hypothesis we intend to verify in this study.

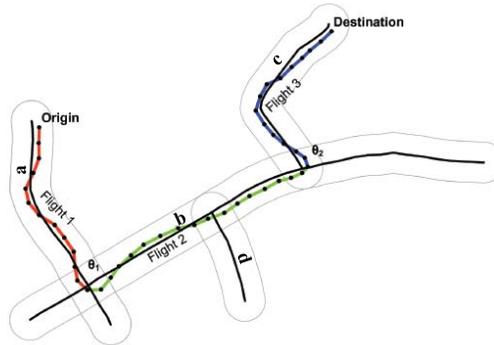

Figure 1: (color online) Illustration of a synthetic trail from the origin to the destination and its flights

The fundamental idea behind this hypothesis can be further addressed by how intensively a street is used by vehicles. By street, we refer to it as a set of consecutive street segments that can be naturally perceived as a street with a good continuity – the natural street (Thomson 2003; Jiang and Liu 2007) or the self-organized natural street (Jiang, Zhao and Yin 2008). Suppose that every vehicle in a street network is equipped with a global positioning system (GPS) receiver, recording a signal every 10 seconds of the vehicle's position ($x_i, y_i, t_i$) along the street. How intensive a street is used by vehicles can be approximated by the number of recorded positions along the street. Note that the number of positions should be adjusted by the speed effect, i.e., the faster the speed, the fewer number of positions. The number of recorded positions over a sufficiently long time period (e.g., one day for daily traffic) in a street network reflects the intensity of street use or traffic distribution on individual streets. The traffic intensity, or traffic distribution, constitutes a human mobility pattern in the street network. Now, let us assume something different. Suppose that we set up the same number of random walkers and let them move around arbitrarily (see the above behaviour description) in the same street network. Would the random walkers be similar to human beings in forming some mobility patterns? Our primary answer to this question is yes. To put it differently, some mobility pattern formed by the random walkers is surprisingly the same as the one by human beings.

This paper aims to provide statistical evidence to support our answer to the above stated question, and test the hypothesis of the similarity between human mobility pattern and that of random walkers. Although human mobility has been studied using massive trajectories of dollar bills (Brockmann, Hufnage and Geisel 2006), and mobile phones (Gonzalez, Hidalgo and Barabási 2008), as well as the more accurate GPS data capturing trajectories of human movement (Rhee et al. 2008), all the studies without an exception adopt a Euclidean space, assuming human mobility is not constrained by the underlying street network. Furthermore, none of the studies attempted to illustrate a possible mechanism as to why human mobility demonstrates the Lévy flight behaviour. In this regard, our study is the first attempt, to the best of our knowledge, which adopts a network constrained space in illustrating human mobility patterns.

The remainder of this paper is organized as follows. Section 2 presents the data sources for the study. Section 3 introduces the methods for detecting a power law or Lévy flights using movement data. Section 4 examines in details the statistics of observed human mobility such as power law and Lévy flights. Section 5 further simulates random walkers and illustrates their overall mobility properties or patterns. In section 6, we try to uncover a possible mechanism behind the power law distribution and Lévy flight behaviour, and further compare the observed human mobility and simulated random walk movement. Finally, section 7 summarizes the study and points to future work.

**2. Data**
Three data sets are involved in this study: GPS data of 50 taxicabs' positions obtained automatically by GPS receivers, anonymized customer data as to when customers are picked up and dropped off, and the underlying street network. The data were collected from four cities or towns in the middle of Sweden: Gävle, Sandviken, Storvik, and Hofors (Figure 2a). The GPS data are massive, a GPS signal is captured every 10 seconds during a six-month period (October 2007, January – May 2008) for every one of the 50 cabs, capturing 59 983 958 positions. The customer data cover the same time period, recording anonymously a total of 166 679 customers' information as to when they were picked up and dropped off. The GPS data are represented by a quadruple $(cabID, x_i, y_i, t_i)$, recording the cab's position $(x_i, y_i)$ at time $t_i$, while the customers data are represented by a triple $(cabID, t_p, t_d)$, representing pick-up time $t_p$ and drop-off time $t_d$. Apparently, the association between a customer and a cab is known through the associated cab and the association between $(t_p, t_d)$ and $(x_i, y_i, t_i)$. The street network covers an area of 1 253 square kilometers, with in total 2 445 kilometers in length (Figure 2b). Based on the street network, we generate 4 056 natural streets from a total of 10 439 street segments.

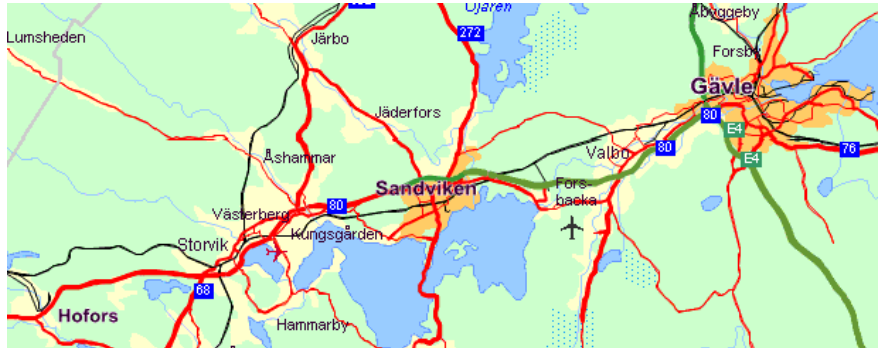
(a)

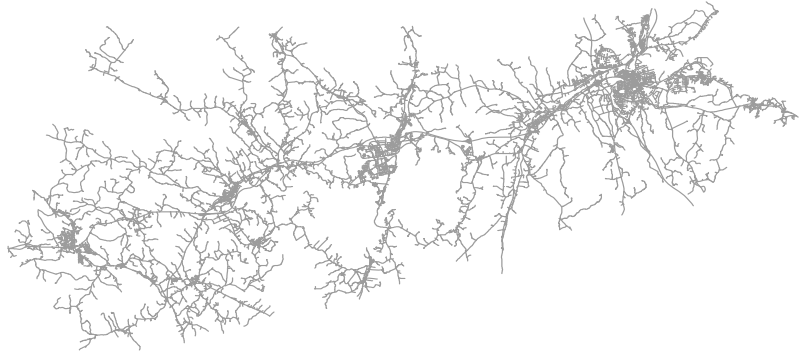

(b)

Figure 2: (colour online) Maps (a) the taxi service area (which is copyrighted by eniro.se), and (b) street network covering the four cities

Both the GPS data and the customer data were used to extract trails. A trail is the GPS trajectory, representing a movement from where one is picked up to where one is dropped off. We derived 72 688 trails from the six-month data sets. Note the difference between the number of trails (72 688) and that of customers (166 679) as some invalid trails were not counted in our statistics. For example, trails lasting more than two hours are considered to be invalid. This is because in a normal condition no one would take a taxi for such a long time. The extracted trails represent a fairly true picture of human mobility, or an overall traffic distribution in the street network. Associated with trails are flights. As an example, Figure 1 illustrates a synthetic trail passing over three different streets *a*, *b* and *c*, and thus it consists of three flights. Flights have two parameters: flight length and turning angle. The turning angle is in fact the deflection angle between the corresponding adjacent street segments. For example, the turning angle between flight 1 and flight 2 is the deflection angle between street *a* and street *b*. We extracted 578 585 flights from the 72 688 trails for the investigation of statistical properties.

**3. Methods for detecting Lévy flights: fitting power laws to human movement data**
Developed by the French mathematician Paul Lévy (1886 – 1971), a Lévy flight model is to extend the famous Brownian motion to be a more general random walk model (Klafter, Shlesinger, and Zumofen 1996). A random walk model is a mathematical formalization of the intuitive idea of taking successive jumps or flights, each in a random direction. A Lévy flight is a particular random walk model that involves two different distributions: a uniform distribution for the turning angle ($\theta_i$) on the one hand, and a power law distribution for the jump length ($\ell_i$) on the other, i.e.,

$$P(\ell_i) \sim \ell_i^{-\alpha} \qquad [1]$$

with $1 < \alpha < 3$.

Herein, the jump length lacks an average scale ($\bar{\ell}$), hence it is often called scale free (Barabási and Albert 1999). It is important to point out that the Lévy stable distribution of the sums of jump length converges to a Gaussian distribution when the Lévy exponent is set to >= 3, indicating a Brownian motion. More significantly, Lévy flights differ from Brownian motions in movement patterns. The jump length of Brownian motions exhibits a Gaussian distribution, while that of Lévy flights possesses a power law distribution (Figure 3).

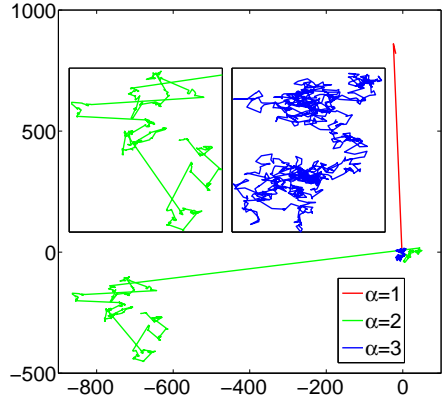

Figure 3: (colour online) Illustration of Lévy flights with different exponent $\alpha$
(NOTE: $\alpha \to 1$, straight lines red; $\alpha \geq 3$, Brownian motion clusters (zoomed inset in blue); $1 < \alpha < 3$, Lévy flights being scale invariant (zoomed inset in green))

To detect a Lévy flight is no more than to detect a power law behavior of the jump length and further estimate the exponent $\alpha$ to see whether it is within the range $1 < \alpha < 3$. This is commonly achieved by plotting the histogram of jump length in logarithmic scales. If the histogram closely follows a straight line, we can claim that a power law exists. This can be seen clearly from the logarithmic transformation of equation [1], i.e.,

$$\log P(\ell_i) = -\alpha \log(\ell_i) + c \qquad [2]$$

where $\alpha$ is the Lévy exponent, and $c$ a constant.

However, the above commonly used method suffers from errors in the logarithmic tail of the distribution (e.g., Goldstein, Morris and Yen 2004; Newman 2005; Sims, Righton and Pitchfford 2007). The end of the logarithmic tail looks messy because each bin only has very few samples in it. One possible solution is to use varying widths of bins in the histogram, with each bin $k$ being increased by $2^k$, to achieve a more homogeneous number of samples per bin. This can help to reduce errors in the tail. This method can be further refined, i.e., by using the frequency per logarithmic bin normalized by dividing by bin width and total number of jumps to get the probability density (e.g., Newman 2005; Viswanathan et al. 1999). This first solution is actually based on the probability density function (PDF). Another solution, probably a much better one, is simply to use the cumulative density function (CDF), i.e., a plot of the probability that the jump length is greater than or equal to a certain value. This form of the power law distribution is sometimes called Zipf's law or Pareto distribution.

The above mentioned alternative solutions do not terminate here if our goal is to obtain the power law exponent. Drawing a conclusion on Lévy flight relies on an accurate estimate of the power law or the Lévy exponent $\alpha$. This is usually done by using the least-squares fit, but it is known to introduce systematic bias (Goldstein, Morris and Yen 2004). Because of the fact that a power law distribution is likely to be confused with other heavy tail distributions, such as the lognormal distribution (Limpert, Stahel and Abbt 2001) and the stretched exponential distribution (Laherrere and Sornette 1998), it is very tricky to draw a power law hypothesis. Many researchers (e.g., Goldstein, Morris and Yen 2004; Newman 2005; Clauset, Shalizi and Newman 2007) have suggested more reliable methods, based on maximum likelihood methods and the Kolmogorov-Smirnov (KS) test, for identifying and quantifying power-law distributions. Their methods can be used not only to fit a power law to data or part of the data, but also for assessing how good the fit is, in comparison with other heavy tailed distributions. The estimated exponent is given by,

$$\alpha = 1 + n \left[ \sum_{i=1}^{n} \ln \frac{\ell_i}{\ell_{\min}} \right]^{-1} \qquad [3]$$

where $\alpha$ denotes the estimated exponent, and $\ell_{\min}$ is the smallest jump length for which the power law holds.

A modified KS test suggested by Clauset, Shalizi and Newman (2007) is adopted in this study to assess the goodness of fit, i.e., how good human movement data fit a power law distribution. A fundamental idea is the maximum distance ($\delta$) between the CDFs of the data and the fitted model:

$$\delta = \max_{x \geq x_{\min}} |f(\ell) - g(\ell)| \qquad [4]$$

Where $f(\ell)$ is the CDF of the synthetic data with a value of at least $\ell_{\min}$, and $g(\ell)$ is the CDF for the power law model that best fits the data while $\ell \geq \ell_{\min}$.

With the fitted model $g(\ell)$, we generate 1000 synthetic datasets that follow a perfect power law above $\ell_{\min}$ but have the same non-power-law distribution as the observed data below, and recalculate the maximum distance between $f(\ell)$ and the fitted model, i.e., $\delta_i (i = 1, 2, ... 1000)$. A goodness of fit index p-value is defined by

$$p = \frac{\text{the number of } \delta_i \text{ whose values are greater than } \delta}{1000} \qquad [5]$$

The p-value indicates to what extent the data fit the model. The larger the p value, the more significant is the model, and p values greater than 0.05 are considered to be acceptable for a goodness of fit. However, some literature suggests a smaller threshold around 0.01 (e.g., Gonzalez, Hidalgo and Barabási 2008).

Although remaining controversial and requiring further and closer scientific scrutiny (Edwards et al. 2007), Lévy flight behaviour has been found in a variety of animal mobility and dispersals (Viswanathan et al. 1999; Viswanathan, Raposo and Luz 2008; Sims et al. 2008) and it becomes an appealing random walk model. Interested readers should refer to Viswanathan, Raposo and Luz (2008) and references therein for more details. The study by Viswanathan et al. (1999) and many follow-up studies (e.g., Sims et al. 2008) have shown that Lévy flights can theoretically increase the success rate of a random search for a diverse variety of animals. Recently, human mobility has also been found to follow the random walk model (Brockmann, Hufnage and Geisel 2006, Rhee et al. 2008), although a continuous Euclidean space is assumed in the studies. In what follows, we adopt a network constrained discrete space, and aim to illustrate human mobility patterns from the perspectives of trails and flights.

**4. Observation of human mobility**

**4.1 Power law fit for trails**
Applying the above methods to the trail length, we found two very striking power laws, respectively, for the trails between 3 km and 23 km, and for the trails greater than 23 km, i.e., $P(\ell_T) \propto \ell_T^{-2.5}$ ($3 \leq \ell_T \leq 23$) and $P(\ell_T) \propto \ell_T^{-4.6}$ ($\ell_T > 23$), where $\ell_T$ denotes trail length (Figure 4). This finding resembles the human behaviour observed by Barabási (2005) on human dynamics. It is also observed that 97% of trails are in the range 3-23 km, while only 3% are greater than 23 km. Table 1 lists the exponent and p value for the goodness of fit. The two ranges (3-23 km, and >23 km) of the trails correspond, respectively, to intra-city and inter-city movements, implying that for both intra-city and inter-city mobility each shows a scaling property. We can note that the exponent for the inter-city mobility trails is too big to make any sense, and it may be simply considered to be a cut-off.

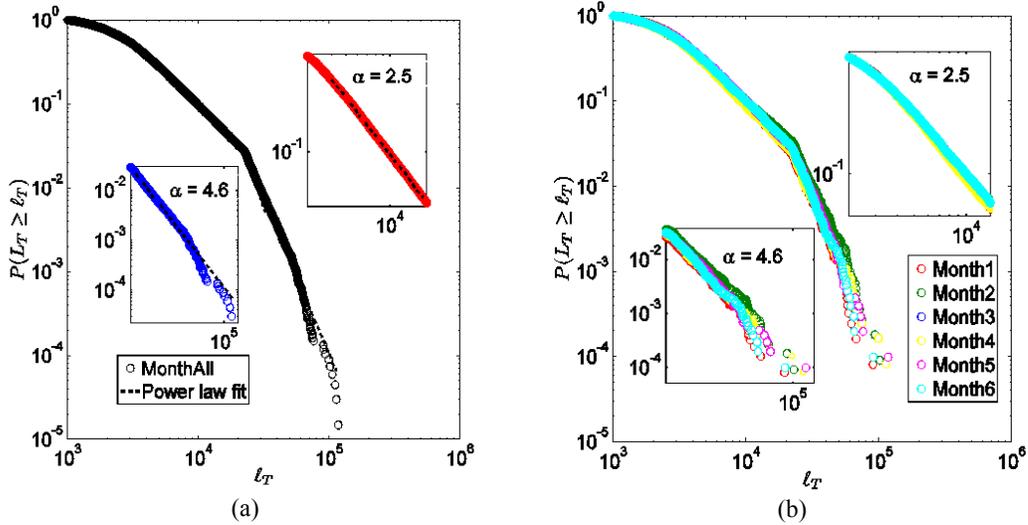

Figure 4: (colour online) (a) Bipartite power law distribution for the trails of 3-23 km and >23 km for all six months together, (b) The same power law distribution plotted month by month for checking consistency

Table 1: Exponent $\alpha$ and p value for both intra- and inter-city trails (support of power law behaviour)

|  | Intra-city trails | | Inter-city trails | |
| --- | --- | --- | --- | --- |
|  | α | p | α | p |
| Month1 | 2.69 | 0.02 | 4.73 | 0.20 |
| Month2 | 2.45 | 0.04 | 4.44 | 0.85 |
| Month3 | 2.48 | 0.49 | 4.63 | 0.03 |
| Month4 | 2.57 | 0.22 | 4.68 | 0.15 |
| Month5 | 2.54 | 0.16 | 4.72 | 0.44 |
| Month6 | 2.52 | 0.09 | 5.05 | 0.51 |
| MonthAll | 2.54 | 0.12 | 4.64 | 0.51 |

**4.2 Power law fit for observed flights**

A flight represents part of a movement along a persistent direction in a continuous space. In the discrete space of a street network, we define a flight as part of a movement along one particular street, which could be linear or curved (Figure 1). With the extracted flights, we find that the flight length fit a power law with an exponential cutoff, i.e., $P(\ell_F) \propto \ell_F^{-2.3} \cdot e^{-\lambda \ell_F}$ ($\lambda = 0.00004$), better than any alternatives such as log-normal, exponential, or stretched exponential cut-off (the details on the goodness of fit test can be found in Table 2). The power law exponent $\alpha$ is between 1 and 3, indicating a Lévy flight behavior. This finding conforms to previous studies on human mobility patterns (Brockmann, Hufnage and Geisel 2006). On the other hand, the turning angle is found to exhibit a very sharply peaked bimodal distribution (Figure 6). This distribution clearly deviates from the definition of Levy flights given in section 3. This deviation is probably due to the fact that human mobility occurred in a network constrained space, while the initial definition of Levy flights is given in a continuous Euclidean space. It should be noted that there is a little bump in Figure 5. This is due to the fact that the number of flights greater than 9 km makes up a very small portion. The same observation can be made regarding the simulated flights, but not as obvious as with the observed flights.

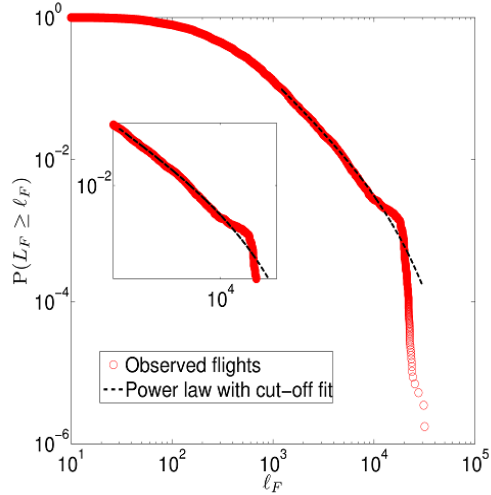

Figure 5: Observed flights showing a Lévy flight behavior

Table 2: Tests of Lévy flight behaviour with observed flights (support of power law with cut-off)
(LR = likelihood ratio, for more details about it, refer to Clauset, Shalizi and Newman (2007))

| Flight | lognormal | | exponential | | stretched exp. | | power law with cutoff | | exponent $\alpha$ | rate $\lambda$ |
|---|---|---|---|---|---|---|---|---|---|---|
| | LR | p | LR | p | LR | p | LR | p | | |
| Month1 | -33.2 | 0.0 | 1343.5 | 0.0 | -35.3 | 0.0 | -45.4 | 0.0 | 2.4 | 0.00005 |
| Month2 | -12.4 | 0.0 | 1581.5 | 0.0 | -13.6 | 0.0 | -27.4 | 0.0 | 2.4 | 0.00003 |
| Month3 | -16.7 | 0.0 | 1503.4 | 0.0 | -17.9 | 0.0 | -28.8 | 0.0 | 2.4 | 0.00003 |
| Month4 | -34.2 | 0.0 | 1303.8 | 0.0 | -36.8 | 0.0 | -49.8 | 0.0 | 2.3 | 0.00005 |
| Month5 | -51.0 | 0.0 | 1056.8 | 0.0 | -54.1 | 0.0 | -64.2 | 0.0 | 2.2 | 0.00006 |
| Month6 | -51.0 | 0.0 | 1152.1 | 0.0 | -53.3 | 0.0 | -60.9 | 0.0 | 2.2 | 0.00006 |
| MonthAll | -181.9 | 0.0 | 7882.7 | 0.0 | -194.1 | 0.0 | -260.5 | 0.0 | 2.3 | 0.00004 |

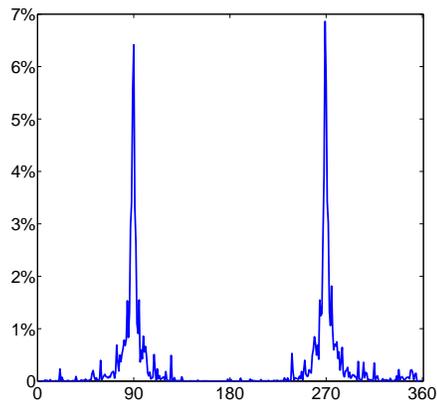

(a)

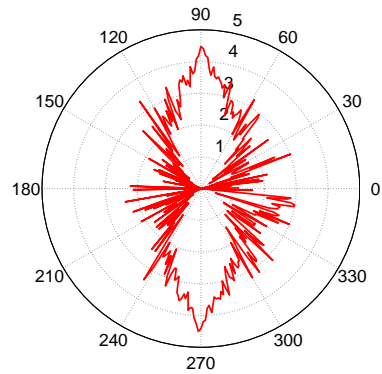

(b)

Figure 6: Distribution of the turning angle with the observed flights. (a) Histogram in the range of 0-360, each subrange (0-180, 180-360) showing a very sharply peaked bimodal distribution, (b) The same histogram in a polar plot with logarithm for the frequency

**4.3 Power law fit for observed traffic (90/10)**
The number of positions recorded along the individual streets indicates to a certain degree the overall traffic distribution. However, this does not take into consideration of the speed effect, i.e., the faster the speed, the fewer number of positions. In order to remove this speed effect, we made an adjustment using the equation:

$$F = \# \cdot \bar{v} \qquad [7]$$

where $\#$ is the number of positions recorded in an individual street, and $\bar{v}$ is the average speed of a cab along the street.

For instance, three different cabs with speeds of 18, 7.2 and 3.6 km/h drive along a 100 meters long street, and the numbers of positions recorded would be 2, 5, and 10, respectively, given that the positions are recorded every 10 seconds. After this speed adjustment, every cab equally contributes to the traffic distribution, no matter what speed each holds.

We found a power law behavior in the observed traffic. This implies that a minority of streets account for a majority of the traffic, i.e., the top 20% of streets account for 80% of traffic (Jiang 2008). With the current study, It is found that the top 10% of streets account for over 90% of traffic (Figure 7, Table 3).

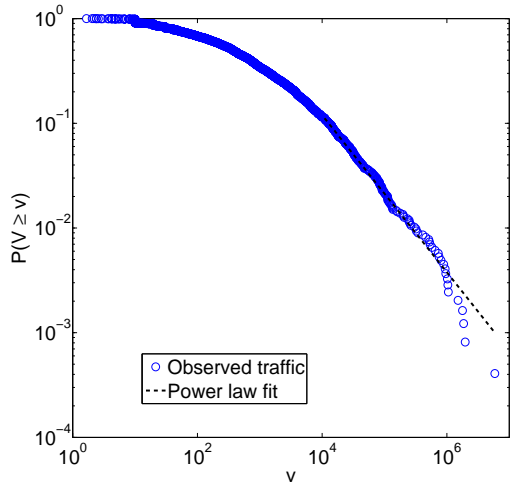

Figure 7: Power law distribution of observed traffic

Table 3: Tests of power law behaviour of observed traffic assigned to individual streets (support of power law behaviour)

|          | α    | p    |
|----------|------|------|
| Month1   | 1.72 | 0.15 |
| Month2   | 1.76 | 0.39 |
| Month3   | 1.75 | 0.29 |
| Month4   | 1.75 | 0.29 |
| Month5   | 1.74 | 0.14 |
| Month6   | 1.78 | 0.28 |
| MonthAll | 1.75 | 0.52 |

## 5. Simulation of random walkers' mobility

### 5.1 Simulation settings
We simulate the mobility of numerous random walkers based on the idea of agent-based simulations, in order to examine whether or not the random walkers can reproduce the observed human mobility patterns. The simulation is carried out from both topological and geometric perspectives respectively (see Appendix A). We adopt two types of random walkers, inspired by the PageRank algorithm (Page and Brin 1998) and its modified version, the weighted PageRank algorithm (Xing and Ghorbani 2004), hence they are referred to as PR-walkers and WPR-walkers. The only difference between the two types of random walkers is whether or not priority is given in choosing the next connected streets. In other words, the PR random walkers are completely random, whereas the WPR ones are less so due to a higher priority given to the most connected streets. Both the random walkers have the damping factor set to 1.0, according to the previous study (Jiang 2008).

The next question is how many ($N_s$) random walkers are used and how long ($T_s$) the simulation will last, as both variables have a significant effect on the number of simulated positions. For example, when $T_s$ is too short, or $N_s$ is too few, most of the streets have no chance to be visited, so the simulation result would be biased. Therefore, the simulation must reach a saturated state before the simulated traffic can be compared against the observed human traffic. The saturated state we set is the time when the frequency of visiting individual streets is highly correlated to the calculated PR or PWR scores. In our simulation, the correlation coefficient R square value is 0.92. This is based on the fact that the probability of a random surfer visiting a web page can be accounted by its PageRank score. Note that the frequency of visiting individual streets and the traffic flow of individual streets are two different concepts, although they are closely related. The former is defined from the topological perspective, and it is counted only once when a connected street is chosen or visited, no matter how long the walkers move along it. However, the latter (traffic flow) is defined from the geometric perspective, and it is accounted for by how long the walkers move along it. To reach the condition of the saturated state, we set either more walkers ($\wedge N_s$) with less time ($\vee T_s$), or fewer walks ($\vee N_s$) with more time ($\wedge T_s$), both having the same effect.

We set up 500 random walkers of each type to move at a persistent speed of 18 km/h for $5 \times 10^6$ seconds: (1) each random walker arbitrarily chooses one of the observed O/D as its origin, and then follows the random behavior, (2) how long each random walker shall walk for is determined by a random generator, which generates a series of travel times ($T_m$) for all the walkers. The travel time ($T_m$) follows a power law distribution with an exponent of 2.5. The reason why we chose this particular exponent is based on the observation of the trail length, the majority of which (97%) exhibit a power law with an exponent of 2.5. The speed of 18km/h reflects the average speed of the 50 taxicabs.

### 5.2 Power law fit for simulated flights
Simulated flights demonstrate Levy flight behaviour, but more precisely, a power law with cut-off (Figure 8, Table 4). We are unable to do the test for all the six-month flights together because of the huge size, but the individual one-month flights do suggest a consistent result, i.e., a support of a power law with cut-off. On the other hand, the turning angle is found to exhibit a very sharply peaked bimodal distribution (Figure 9 and 10). A slight difference can be observed in the distribution of turning angles with the polar plots, i.e., the one for PR-walkers is a bit fatter than the one for WPR-walkers. Also, two peaks are a bit different; in human walkers, right turns are more common than left turns, while it is the opposite in random walkers. As to the reason, we suspect that it is to do with the randomness of random walkers, and the purposiveness of human movement.

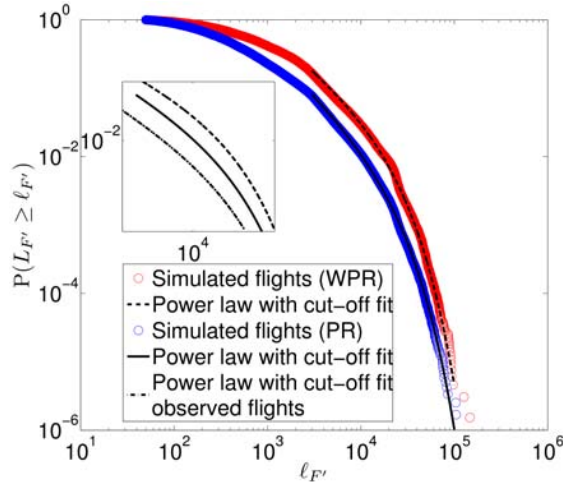

Figure 8: Simulated flights approximately exhibit the same Lévy behavior, i.e., $P(\ell_{F'}) \propto \ell_{F'}^{-2.4} \cdot e^{-\lambda \ell_{F'}}$ ($\lambda = 0.00003$ for WPR-walkers, and $\lambda = 0.00004$ for PR-walkers), in comparison with observed ones (inset)

Table 4: Tests of Lévy flight behaviour with simulated flights (support of power law with cut-off)
(LR = likelihood ratio, for more detail about it, refer to Clauset, Shalizi and Newman 2007)

|  | lognormal | | exponential | | stretched exponential | | power law with cutoff | | Exponent | rate |
| --- | --- | --- | --- | --- | --- | --- | --- | --- | --- | --- |
|  | LR | p | LR | p | LR | p | LR | p | α | λ |
| WPR-walkers | | | | | | | | | | |
| Month1 | -915.91 | 0.00 | 12967 | 0 | -1018 | 0.00 | -1328.7 | 0 | 2.0847 | 0.00003 |
| Month2 | -838.44 | 0.00 | 13647 | 0 | -932.9 | 0.00 | -1242.7 | 0 | 2.12556 | 0.00003 |
| Month3 | -872.80 | 0.00 | 13602 | 0 | -976.8 | 0.00 | -1317 | 0 | 2.096245 | 0.00003 |
| Month4 | -894.62 | 0.00 | 13477 | 0 | -997.3 | 0.00 | -1319.5 | 0 | 2.094951 | 0.00003 |
| Month5 | -896.77 | 0.00 | 13011 | 0 | -997.8 | 0.00 | -1310.3 | 0 | 2.089882 | 0.00003 |
| Month6 | -990.32 | 0.00 | 13487 | 0 | -1102 | 0.00 | -1435.5 | 0 | 2.066448 | 0.00003 |
| MonthAll | N/A | N/A | N/A | N/A | N/A | N/A | N/A | N/A | N/A | N/A |
| PR-walkers | | | | | | | | | | |
| Month1 | -548.45 | 0.00 | 7328 | 0 | -600.8 | 0.00 | -754.43 | 0 | 2.274887 | 0.00004 |
| Month2 | -578.23 | 0.00 | 7274.9 | 0 | -634 | 0.00 | -791.21 | 0 | 2.252193 | 0.00004 |
| Month3 | -665.81 | 0.00 | 6991.2 | 0 | -726.6 | 0.00 | -877.13 | 0 | 2.205745 | 0.00004 |
| Month4 | -682.51 | 0.00 | 6903.2 | 0 | -745.1 | 0.00 | -905.56 | 0 | 2.204626 | 0.00004 |
| Month5 | -604.54 | 0.00 | 6850.3 | 0 | -666.1 | 0.00 | -825.92 | 0 | 2.229495 | 0.00004 |
| Month6 | -641.62 | 0.00 | 6891.3 | 0 | -704.9 | 0.00 | -870.96 | 0 | 2.208273 | 0.00004 |
| MonthAll | N/A | N/A | N/A | N/A | N/A | N/A | N/A | N/A | N/A | N/A |

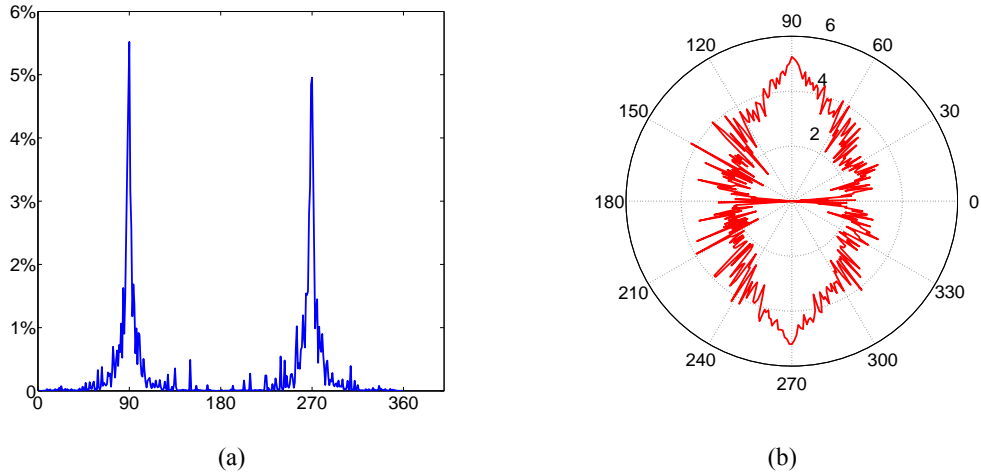

Figure 9: Distribution of the turning angle with the simulated flights (WPR-walkers). (a) Histogram in the range of 0-360, each subrange (0-180, 180-360) showing a very sharply peaked bimodal distribution, (b) The same histogram in polar plot with logarithm for the frequency

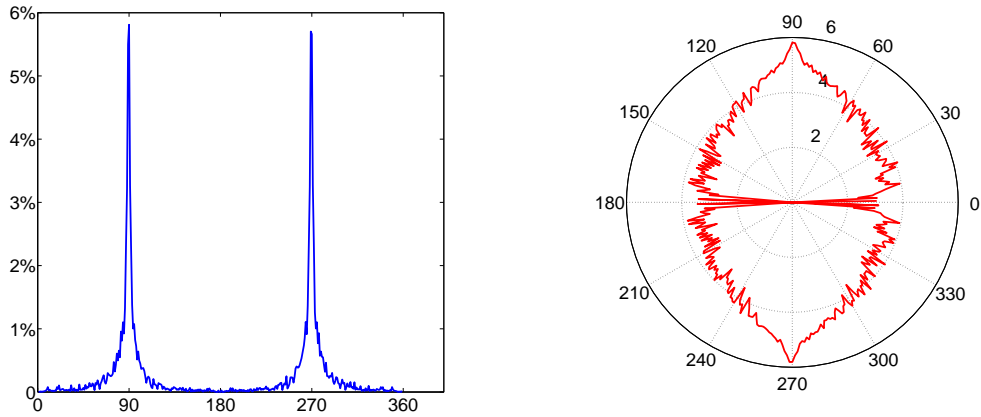

Figure 10: Distribution of the turning angle with the simulated flights (PR-walkers). (a) Histogram in the range of 0-360, each subrange (0-180, 180-360) showing a very sharply peaked bimodal distribution, (b) The same histogram in a polar plot with logarithm for the frequency

**5.3 Power law fit for simulated traffic (90/10)**
Power law behavior is again found in the simulated traffic (Figure 11, Table 5), implying that a minority of streets account for a majority of traffic, i.e., the top 10% of streets account for over 90% of traffic.

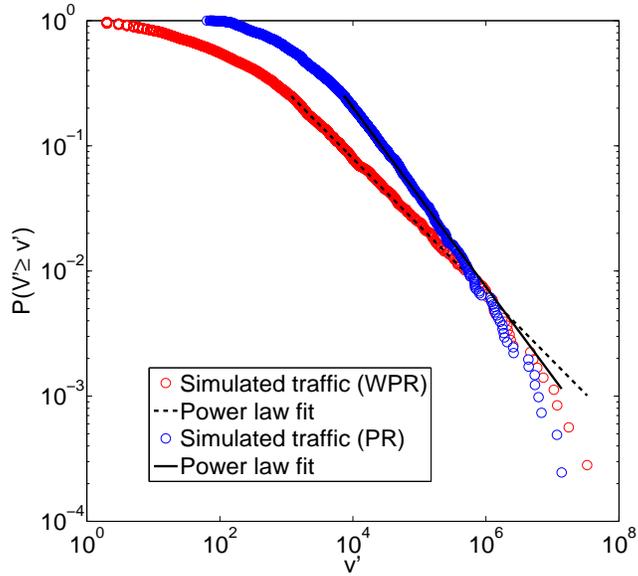

Figure 11: Power law distribution of simulated traffic

Table 5: Tests of power law behaviour with simulated traffic assigned to individual streets (support of power law behaviour)

|          | WPR-walkers | | PR- walkers | |
|----------|------|------|------|------|
|          | α    | p    | α    | p    |
| Month1   | 1.53 | 0.65 | 1.66 | 0.19 |
| Month2   | 1.54 | 0.33 | 1.77 | 0.10 |
| Month3   | 1.53 | 0.59 | 1.79 | 0.19 |
| Month4   | 1.53 | 0.13 | 1.94 | 0.53 |
| Month5   | 1.55 | 0.53 | 1.77 | 0.20 |
| Month6   | 1.53 | 0.34 | 1.79 | 0.54 |
| MonthAll | 1.54 | 0.59 | 1.72 | 0.11 |

**6. Mechanism behind the Levy flight behavior and observed traffic versus simulated traffic**

**6.1 A possible mechanism behind power law behavior of trail length**
To uncover a possible mechanism behind the scaling property, we further examine the distribution of all possible synthetic trails between the origins and destinations (O/D) (Figure 12a). The synthetic trails are assumed to be the shortest paths between the pairs of O/D. The number of O/D is massive, about 25 000 in one month, generating about 312 500 000 synthetic trails using the well-known Dijkstra shortest path algorithm. This number would make the power law fit test too computationally intensive. Thanks to the clustered nature of O/D, and in order to reduce the computational burden, we first adopted the k-means clustering method (Hartigan and Wong 1979) to create 4 000 clusters from 25 000 O/D in a one month period. The gravity centers of the 4000 clusters are considered to be new O/Ds, which are snapped into the street network to derive (4000*4000)/2-4000 = 7 996 000 shortest paths or synthetic trails. The probability density function (PDF) of the synthetic trails are then plotted to assess a power law fit. Although no striking power laws exist with the synthetic trails, the CDF of the synthetic trails in the ranges 3-23km, and >23km does show a tendency for a power law distribution (insets with Figure 12). It is observed that about 80% of O/D are scattered, while 20% of them, indicated in red, are highly clustered, showing a clear scale-invariance and fractal property. It is the scale free nature of the O/D distribution that underlies the power law distribution of trail length.

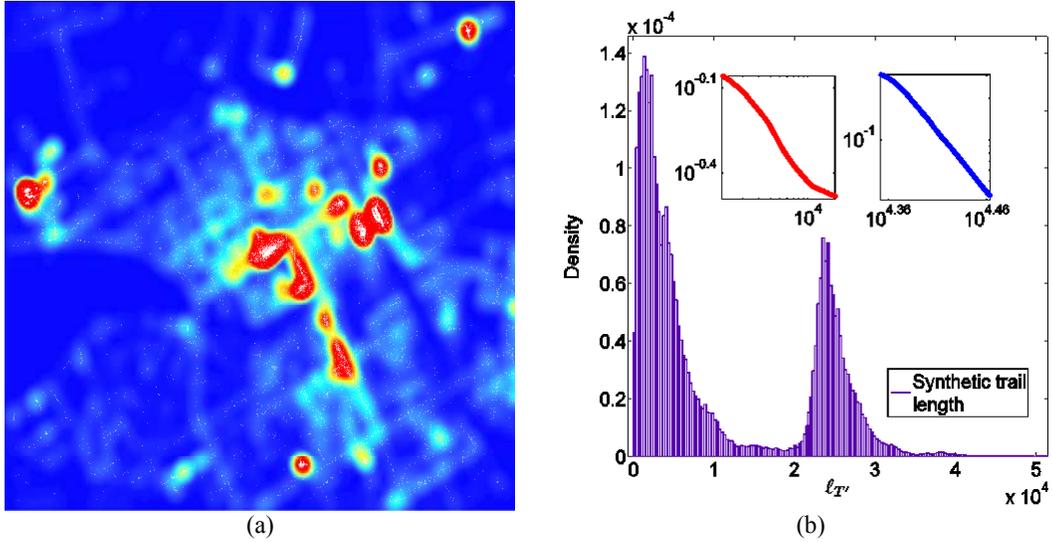

(a)                                                  (b)

Figure12: A possible mechanism behind the power law behavior of trail length: (a) The tiny white spots are O/D during January 2008 at the centre of Gävle city. The density of O/D is visualized in color from the highest (red) to the lowest (blue). The largest red hotspot in the center is the downtown square, and other hotspots are the hospital, the central station, a district center, and the location of the taxi company. (b) The PDF of the 7 996 000 synthetic trails generated from the 4000 clusters, and the CDF of the trails in the ranges of 3-23 km (inset in red), and >23 km (inset in blue), indicate two heavy-tail distributions.

**6.2 A possible mechanism behind Levy flights behavior**
We believe that it is the underlying street network that governs the Lévy flight behavior. To verify this belief, we further examine the topological structure of the street network in terms of the street-street intersections. First we derived 4 056 natural streets based on the every-best-fit join principle and using degree 45 as the threshold. The reason why we adopted this particular principle is because the natural streets formed by this principle tend to be similar in morphological structure to the named streets (Jiang and Claramunt 2004). The generated streets are transformed into a connectivity graph in terms of which street intersects which street, a sort of street topology based on street-street intersections. Both street length and connectivity are assessed to see whether or not they follow a power law distribution. Not surprisingly, the underlying street topology demonstrates a power law distribution with the exponent around 2.5, i.e., $P(c) \propto c^{-2.6}$ for street connectivity, and $P(\ell_S) \propto \ell_S^{-2.4}$ for street length (Figure 13). This statistical regularity can be expressed by the 80/20 principle: 80% of streets are less connected below the average, and 20% of streets are well connected above the average (Jiang 2007). The recorded taxicabs' positions every 10 seconds are overlapped with the street network to get an overall traffic distribution. We illustrate the fact that a minority of streets account for a majority of traffic (Jiang 2008), but with the current study it is adjusted to be that 10% of well connected streets account for 90% of traffic.

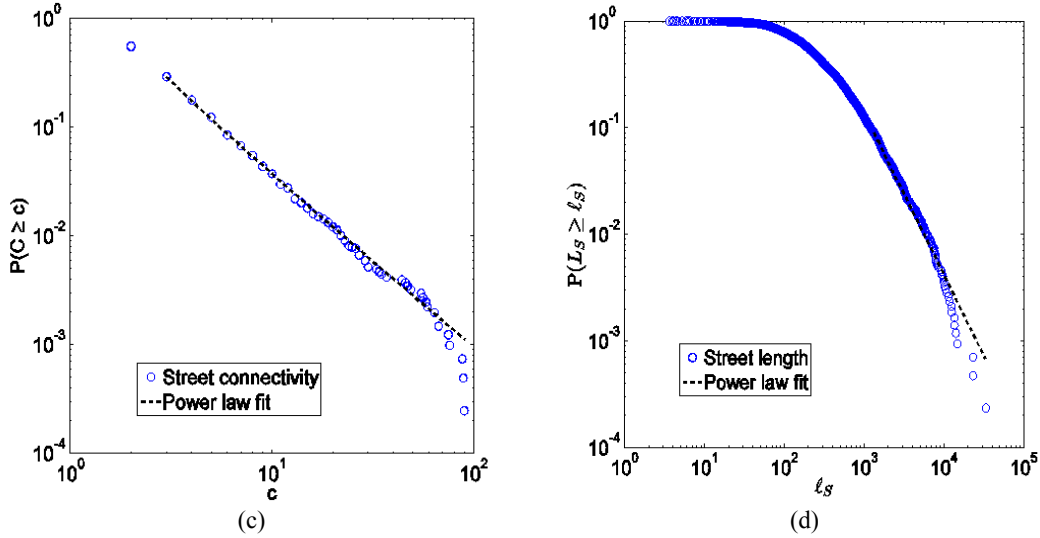

Figure 13: (color online) Scaling of streets (a) The power law distribution of street connectivity, i.e., $P(c) \propto c^{-2.6}$, and (b) The similar power law distribution of street length, i.e., $P(\ell_S) \propto \ell_S^{-2.4}$, where the exponent is slightly less than that of the street connectivity's distribution.

## 6.3 Observed traffic versus simulated traffic

To our surprise, the simulated traffic is highly correlated to the observed one, and the relevant R square value is between 0.83 and 0.87. This finding implies that a significant fraction of traffic can be accounted for by the underlying space (the street network topology and spatial distribution of O/D), and human free will or the purposive nature of mobility has little effect on the traffic distribution.

Table 6: R square values between observed traffic and simulated traffic assigned to individual streets (support of a significant correlation)

|  | WPR-walkers | PR-walkers |
| --- | --- | --- |
| Month1 | 0.85 | 0.85 |
| Month2 | 0.87 | 0.83 |
| Month3 | 0.86 | 0.83 |
| Month4 | 0.85 | 0.85 |
| Month5 | 0.85 | 0.84 |
| Month6 | 0.86 | 0.83 |

## 7. Conclusion and future work

In summary, in this study we investigated human mobility patterns from two different perspectives: trails between O/D and flights between streets. We found scaling properties with both trails and flights. The flights show a Lévy behavior in nature. More importantly, through simulation and experiments, it is illustrated that the scaling properties are attributed to the underlying street network and spatial distribution of O/D. We further conjecture that such a mechanism can be applied to human mobility patterns at a larger scale, e.g., at a country level. It is the fractal property of space itself that governs the scaling properties of human mobility patterns. Herein the space refers to man-made infrastructure such as cities, airports, train stations, and highways that all bear the scale-invariance property that governs the overall human mobility patterns. In this regard, the study by Barthélemy and Flammini (2006) provides an interesting model. Our future work should concentrate on further investigation of geographic space, i.e., how does space in general have an effect on human mobility patterns?

**Acknowledgement**
The research idea was conceived by BJ, data processing, analysis and simulations were done by BJ, JY and SZ and the manuscript was written by BJ. This study is financially supported by research grants from the Hong Kong Polytechnic University and the Swedish Research Council FORMAS. We would like to thanks TAXI Stor och Liten for kindly providing the mobility data. The street network data are provided by the city of Gävle. In this connection, Anders Wahlgren and Annelie Hook deserve our special thanks for the data preparation and transferring. JY would like to thank Aaron Clauset and César Hidalgo for their advice on the modified KS-test.

**Appendix A: Algorithm for simulating mobility of random walks**

The algorithm is a detailed implementation of the random movement behavior given at the beginning of the paper. It is important to note a few points on the implementation: (1) each random walker arbitrarily chooses one of the observed O/D as its origin, and then follows the random behavior, (2) how long each random walker shall walk for is determined by a random generator, which generates a series of travel time (Tm) for all the walkers. The time (Tm) follows a power law distribution with an exponent of 2.5. The reason why we choose the exponent range is based on the observation of the trail length, the majority of which (97%) exhibit a power law with the exponent of 2.5.

```
Input: A street-based network shape file
Output: A txt file recording the number of positions in walker's trajectories for each street
Variables: D: damping factor value
           T: simulation period
--------------------------------------------------------------------------------
//The main function of simulation
Sub Simulating()
  Transform the street-based network into a connectivity graph;
  Create N turtles distributed randomly in the OD-located streets;
  Use a power-law (α=2.5) function to randomly create a travel time Tm for each walker;
  t = 0;
  While (t <= T) do
    For Each walker in Walkers
        If the walker's real travel time > Tm Then
          Let walker jump randomly to an OD-located street;
          Use a power-law (α=2.5) function to randomly re-create a travel time Tm for it;
        End If
        Call walker::MoveOn(dt) function to let the walker move to a new location;
    Next
    t = t + dt;
  End while
  //Output
  Output the number of locations for each road to a txt file;
End sub

--------------------------------------------------------------------------------
CLASS: Walker
// Walker walking behavior
Sub Walker::MoveOn(dt)
  If (walker's next street doesn't exist) Then
     Call walker::SearchNextStreet function to get a next street;
     Get the junction between walker's current street and this street;
     If (no such junction) Then
        //Leap to another street
        Locate the walker in the middle of this street;
        Assign this street to the walker's current street;
        Increment the walker-tracked number of current street by one;
        Set the walker's next street to null;
        Return;
     Else
        Assign this street to the walker's next street;
        Compare the spatial relationship between the walker's current position and the
         junction to decide the walker's walking direction;
     End If
  End If

  Use walker's speed, direction and time increment to calculate the next position;

  If (walker reaches the junction between current street and next street) Then
     Locate the walker in the junction;
     Replace the walker's current street with next street;
     Set walker's next street to null;
     Increment the walker-tracked number of current street by one;
  Else
     Locate the walker in the new position;
     Increment the walker-tracked number of current street by one;
  End If
End sub

//Search a next street based on weighted PageRank (WPR) or standard PageRank (PR)
Function Walker::SearchNextStreet(walker's current street) as street
  Get the connected street list A from walker's current street;
```

```
    Generate a random number 0<d<1;
    If (d <= D) Then
        If (PR) Then
            Pick up a street randomly from List A;
            Return this street;
        Else if (WPR) Then
            Pick up a street from List A according to the probability p;
            // P is decided by the weight (w = connectivity / sum of connectivity)
            // this pick up is in contrast to random pick up
            Return this street;
        End If
    Else (d > D) Then
        Select a street randomly;
        Return this street;
    End If
End Function
```